\documentclass[aps,prl,twocolumn,superscriptaddress]{revtex4-1}
\pdfoutput=1

\usepackage{graphicx}
\usepackage{amsmath}
\usepackage{amssymb}
\usepackage{color}
\usepackage{bm}
\usepackage{subfigure}
\usepackage{appendix}

\definecolor{red}{rgb}{0.75,0,0}
\definecolor{blue}{rgb}{0,0,0.75}
\definecolor{green}{rgb}{0,0.5,0}

\newcommand{\unit}[1]{\ensuremath{\, \mathrm{#1}}}
\newcommand{\beq}{\begin{equation}}
\newcommand{\eeq}{\end{equation}}
\newcommand{\beqa}{\begin{eqnarray}}
\newcommand{\eeqa}{\end{eqnarray}}

\begin{document}
\title{Contractile stresses in cohesive cell layers on finite-thickness substrates}

\author{Shiladitya Banerjee}
\affiliation{Department of Physics, Syracuse University, Syracuse New York, 13244-1130, USA}

\author{M. Cristina Marchetti}
\affiliation{Department of Physics, Syracuse University, Syracuse New York, 13244-1130, USA}
\affiliation{Syracuse Biomaterials Institute, Syracuse University, Syracuse New York, 13244-1130, USA}

\date{\today}

\begin{abstract}
Using a minimal model of cells or cohesive cell layers as continuum active elastic media, we examine the effect of substrate thickness and stiffness on traction forces exerted by strongly adhering cells. We obtain a simple  expression for the length scale controlling the spatial variation of stresses in terms of cell and substrate parameters that describes the crossover between the thin and thick substrate limits. Our model is an important step towards a unified theoretical description of the dependence of traction forces on cell or colony size, acto-myosin contractility, substrate depth and stiffness, and strength of focal adhesions, and makes experimentally testable predictions.
\end{abstract}

\maketitle

Many cell functions, such as spreading, growth, differentiation and migration,  are affected by the elastic and geometric properties of the extracellular matrix~\cite{Harris1980}. Considerable effort has been devoted to the study of  cell adhesion to elastic substrates~\cite{Discher2005}. Cells adhere to a substrate via focal adhesion complexes that link the substrate to the actomyosin cytoskeleton, which in turn generates contractile forces that deform soft substrates~\cite{Balaban2001}. The traction forces that the cell exerts on the substrate are regulated by the cell itself in a complex feedback loop controlled by  cell activity and substrate elasticity.

\begin{figure}
\begin{center}
\includegraphics[width=0.45\textwidth]{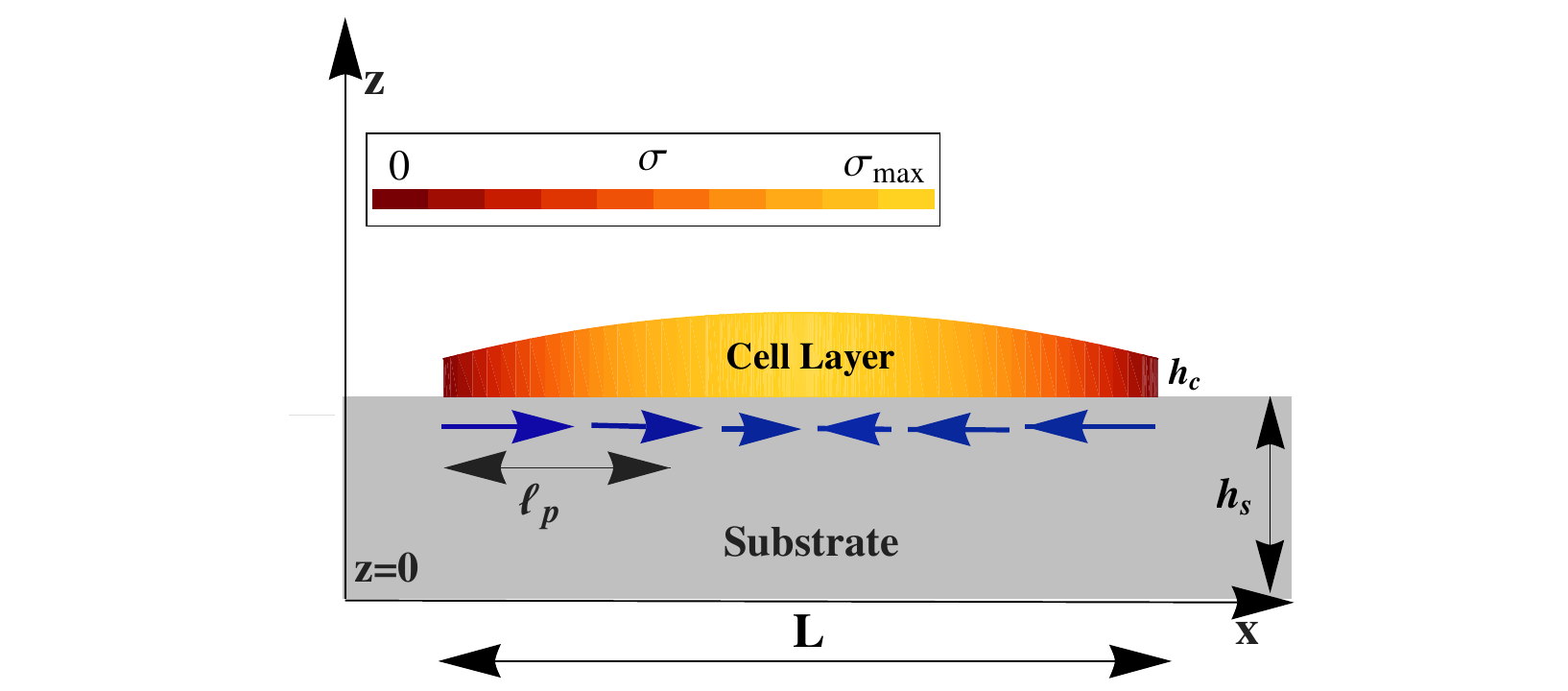}
\end{center}
\caption{Schematic of a cell layer of lateral extent $L$ and thickness $h_c<<L$ adhering to a substrate of thickness $h_s$. The  build-up of contractile stress $\sigma$ in the cell layer is indicated by the color map, while the traction stresses in the substrate are shown as vectors (blue online). The spatial variation of both traction and cellular stresses in the lateral ($x$) direction are characterized by the length  scale $\ell_p$, referred to as the penetration length.}
\label{cell_layer}
\end{figure}

\begin{figure}
\centering
\includegraphics[width=0.5\textwidth]{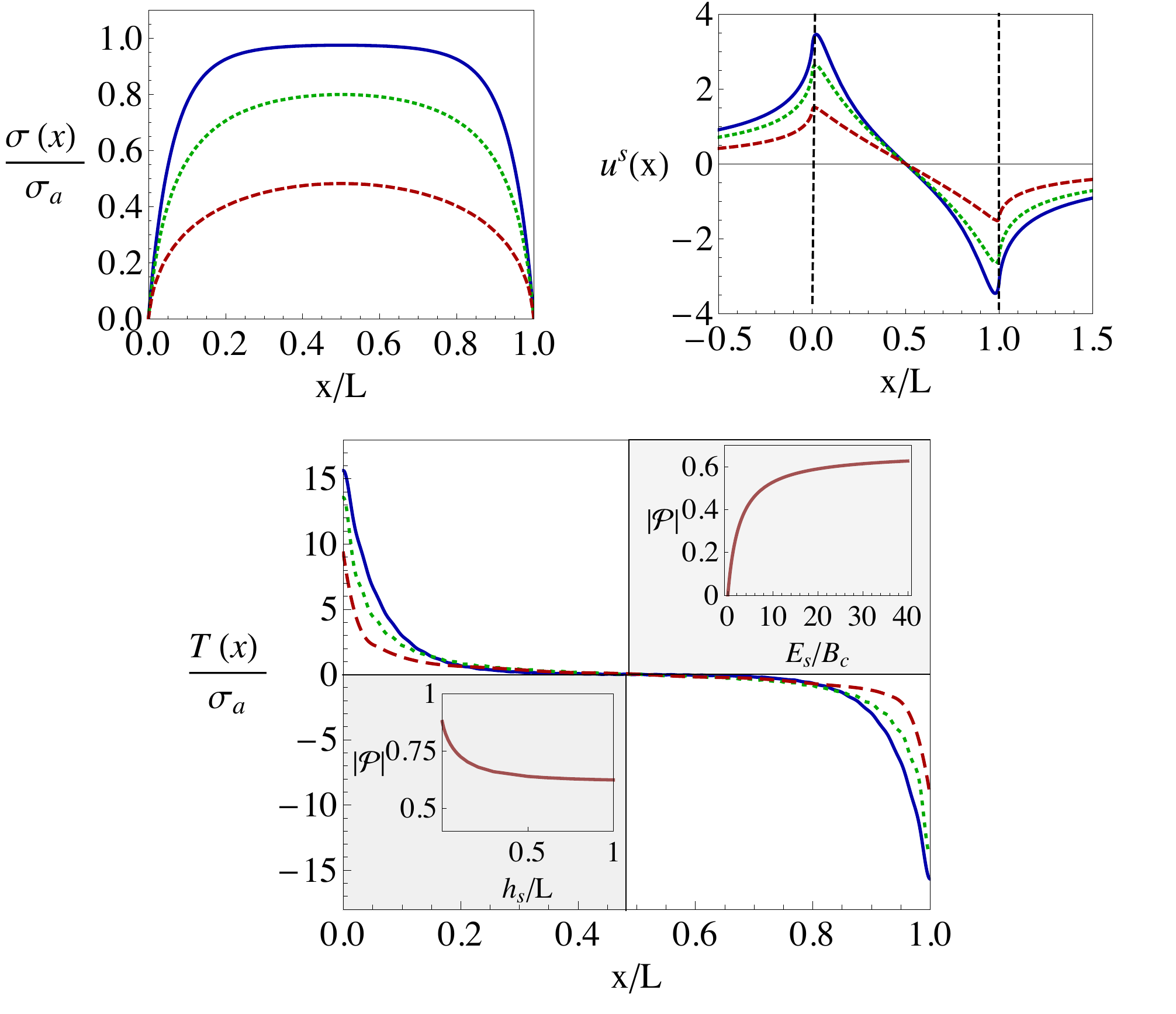}
 \caption{Internal stress $\sigma(x)/\sigma_a$ (top left), substrate displacement $u^s(x)$ (top right) and traction stress $T(x)/\sigma_a$ (bottom) vs position $x$ along the cell layer, for $E_s=500 \unit{Pa}$ (solid, blue), $50 \unit{Pa}$ (dotted, green) and $10 \unit{Pa}$ (dashed, red). The vertical dashed lines in the top right frame denote the cell layer's edges. Inset (bottom right): magnitude of contractile moment $\vert{\cal P}\vert$ vs $E_s/B_c$. Inset (bottom left): $\vert{\cal P}\vert$ as a function of substrate thickness for $E_s=10 \unit{Pa}$. Other parameters: $B_c=1 \unit{kPa}$, $h_c=0.1 \unit{\mu m}$, $\ell_a=6.3 \unit{\mu m}$, $L=100 \unit{\mu m}$, $\nu_s=0.4$.}
 \label{subs}
 \end{figure}

Two powerful experimental techniques have been developed to measure forces  by cells on substrates: traction force microscopy, used to probe cell adhesion to continuous substrates~\cite{Dembo1999,Butler2002}, and the imaging of cell-induced bending of microfabricated pillar arrays~\cite{Tan2003}. These two techniques have also been recently combined~\cite{Polio2012}. These experiments have yielded new insight on substrate rigidity sensing and have opened up new questions on the physics of individual and collective cell adhesion: What controls the length scale that governs the penetration of traction forces?
What is the relative role of active cellular contractility and cell-cell-interaction in controlling the emergent  response of cell layers?
In this Letter we describe minimal models of individual cells and adhering cell colonies that  reproduce qualitatively several experimental findings.
The traction stresses exerted by cells on substrates are extracted directly from measurements of micropillar displacements  or inferred from the displacements of fiducial markers embedded in a continuum substrate.
It is found that traction stresses by isolated fibroblasts and epithelial cells on pillar arrays  are localized near the cell edge, while contractile stresses (referred to below as cellular stresses) built-up inside the cellular material is largest near the cell center~\cite{Dembo1999,Ghibaudo2008}, as shown schematically in Fig.~\ref{cell_layer}. This behavior, also observed in adherent cell sheets and in migrating cell colonies~\cite{Saez2010,Mertz2012,Trepat2009}, is  predicted by our model.
Further, both substrate thickness and stiffness affect cellular and traction stresses~\cite{Lin2010}. The magnitude of the traction stress  increases with substrate stiffness, saturating at large stiffness~\cite{Ghibaudo2008}, and it decreases sharply with substrate thickness, indicating that cell colonies on thick substrates only probe a portion of substrate of effective depth comparable to the lateral extent of the cell colony~\cite{Sen2009}. Both trends are reproduced by our model (Fig.~\ref{subs}).
%

Our model builds on recent work~\cite{Edwards2011,Banerjee2011} describing the cell or cell layer as a contractile elastic medium, with local elastic response of the substrate (as appropriate for micropillar arrays or very thin substrates). In contrast, here we consider substrates of finite thickness where the nonlocality of the elastic response must be included.  While previous studies have analyzed the deformations of finite-thickness substrates due to point traction forces on their surface~\cite{Merkel2007,Maloney2008}, our work considers the inhomogeneous traction due to an extended contractile cell layer. A central result for our work is the expression for the scaling parameter referred to as the lateral penetration length $\ell_p$ (Fig.~\ref{cell_layer}). This length scale characterizes the in-plane spatial variations of both
adhesion-induced traction stresses on the substrate and cellular stresses within the cell layer in terms of cell and substrate elastic and geometrical properties. Our model also quantifies the experimentally-observed role of substrate thickness $h_s$ in controlling the mechanical response of adhering cell layers~\cite{Lin2010}. If $h_s$ is small compared to the lateral extent $L$ of the cell sheet, the substrate elasticity plays a negligible role in determining  the mechanical response of the cell. This may explain why traction forces exerted by cell colonies with $L\gg h_s$ appear insensitive to substrate stiffness~\cite{Trepat2009}.
If, in contrast,  $L\ll h_s$, then
substrate nonlocality controls stress build-up in the cell sheet. This crossover may be observable in large cell colonies on thick substrates.
Finally, the importance of long-range substrate elasticity has also been emphasized in recent models of cells as active dipoles on a soft elastic matrix, where it is crucial in controlling cell adhesion~\cite{Bischofs2004,De2007}. Long-range interfacial elastic stresses coupled with gel thickness have also been shown to have a profound effect on focal adhesion growth~\cite{Nicolas2006} and to  enhance cell polarization~\cite{Bischofs2005,Friedrich2012}. These important effects are not discussed here.

\begin{figure}
\begin{center}
\includegraphics[width=0.45\textwidth]{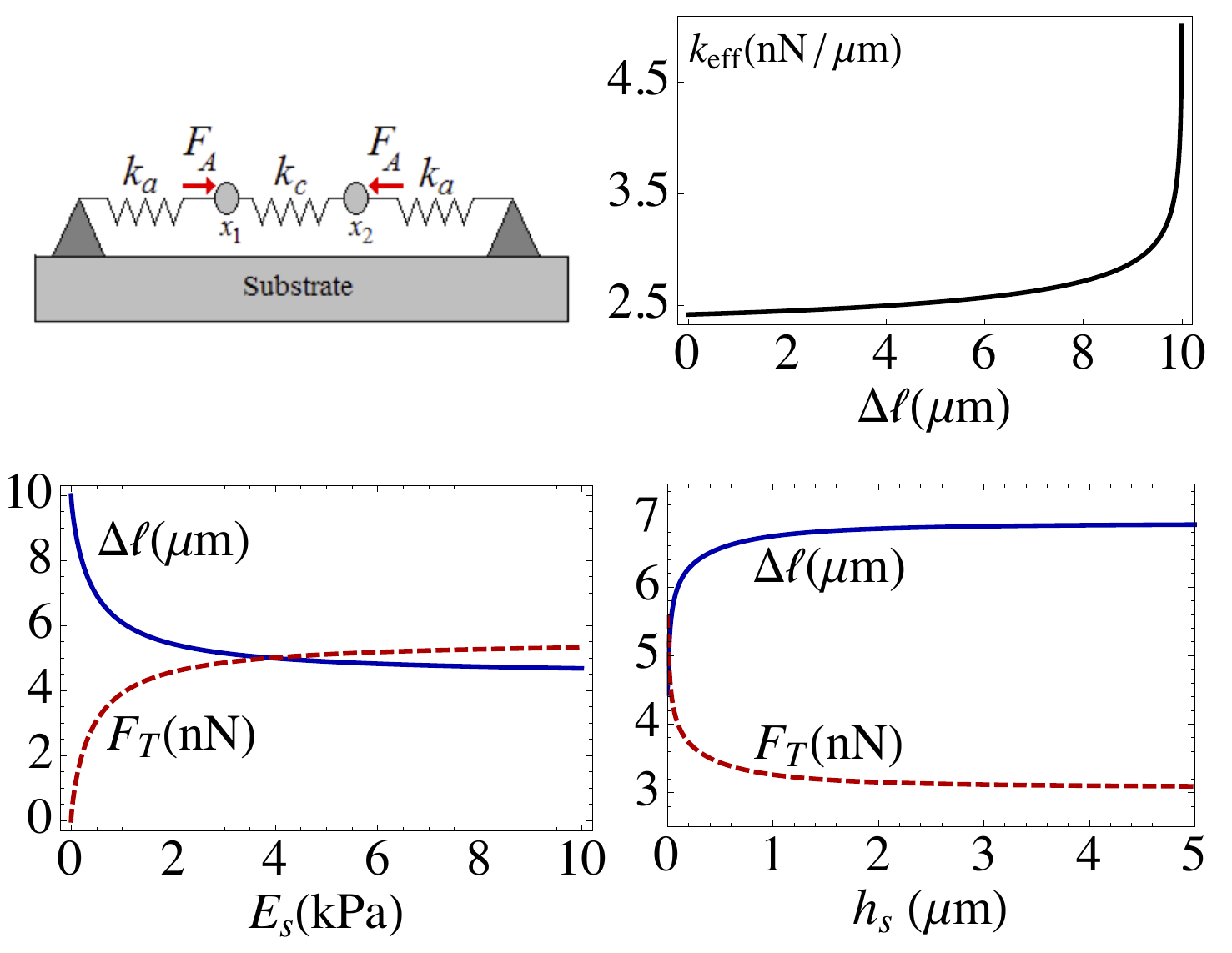}
\end{center}
\caption{Top : Schematic of a contractile cell adhering to a soft substrate (left) and effective spring constant $k_{\text{eff}}$ versus cellular strain $\Delta\ell$, showing strain stiffening (right). Bottom : Cell contraction $\Delta\ell$ (solid blue line) and traction force $F_T$ (red dashed line) vs substrate stiffness (left) for  ${h}_s=10 \unit{\mu m}$  and as a function of substrate thickness (right) for ${E}_s=500 \unit{Pa}$. Other parameters : $F_A=10 \unit{nN}$, $k_c=1 \unit{nN/\mu m}$, $k_a=2.5 \unit{nN/\mu m}$, $E_s=1 \unit{kPa}$, $h_s=10 \unit{\mu m}$, $\ell_{c0}=10 \unit{\mu m}$, $\nu_s=0.4$.}
\label{cs}
\end{figure}

\paragraph{Contractile cell on a soft substrate.}
To illustrate the importance of substrate nonlocality, we first analyze a single cell, modeled as a contractile spring of stiffness $k_c$  and rest length $\ell_{c0}$, adhering to a continuum substrate (described as an elastic continuum  of Young's modulus $E_s$ and Poisson's ration $\nu_s$) via two focal adhesion bonds (linear springs of stiffness $k_a$) located at $x_1$ and $x_2$ (Fig.~\ref{cs}, top left)~\cite{Schwarz2006}. This is motivated by the experimental observation that in adhering cells focal adhesions tend to be localized near the cell periphery~\cite{Wozniak2004}.
For simplicity we consider a one dimensional model, where the cell lies on the $x$ axis and the substrate lies in the $0\leq z\leq h_s$ region of the $xz$ plane.
Contractile acto-myosin fibers connect the focal adhesions and exert active forces of magnitude $F_A$.
Once the cell has fully adhered, the cell-substrate system is in mechanical equilibrium. Force balance at $x_1$ and $x_2$ yields
\begin{subequations}
\begin{gather}
k_a\left[u_1-u^s(x_1)\right]= F_A - k_c(u_1-u_2) \label{u1} \;,\\
k_a\left[u_2-u^s(x_2)\right]=-F_A + k_c(u_1-u_2) \label{u2} \;,
\end{gather}
\label{u}
\end{subequations}
with $u_i$ the displacements of the contact points $x_i$  from their unstretched positions  $x_2^0-x_1^0=\ell_{c0}$, and $u^s(x_i)$ the displacement of the substrate's surface at  $x_i$. All displacements are defined with respect to an initial state where the cell  has length $\ell_{c0}$. The net contraction is then $\Delta \ell=l_{c0}-(x_2-x_1)=u_1-u_2$.
The traction force  by the cell on the  substrate is localized at $x_1$ and $x_2$, yielding a traction force density ${f}_T({x})=F_T\delta({x}-x_1)-F_T\delta({ x}-x_2)$, with $F_T=F_A-k_c\Delta\ell$.
Assuming linear elasticity, the substrate deformation is~\cite{Landau1986},
%
$u^s({x})=\int_{-\infty}^\infty d{x'} G({x}-{x}')f_T({x}')$,
%
where $G(x)$ is the elastic Green's function at $z=h_s$. For a substrate of thickness $h_s$ we use the approximate form~\footnote{To enable a direct comparison between the penetration lengths obtained below and experimentally accessible parameters, $E_s$ is the Young modulus of a three dimensional elastic medium.}
\begin{equation}
\label{G-finite}
G(x)=\frac{2}{\pi \ell_{c0}E_s} K_0\left[\frac{a+\vert x\vert}{h_s(1+\nu_s)}\right]
\end{equation}
derived in the Supplemental Material~\cite{SMprl12}, with $a$ the size of adhesion complexes, providing a short-distance cut-off, and $K_0$ denotes the modified Bessel function of the second kind. We obtain $F_T(\Delta\ell)=\frac{1}{2}k_{\text{eff}}(\Delta\ell)\Delta\ell$, with  $k_{\text{eff}}^{-1}=k_a^{-1}+[G(0)-G(l_{c0}-\Delta\ell)]$ the effective stiffness of the  cell-substrate adhesions.
For $\Delta\ell \ll l_{c0}$, $k_{\text{eff}}$ is independent of $\Delta\ell$ and $F_T$ scales linearly with $\Delta\ell$. Stiffening sets in for $\Delta\ell>\ell_{c0}\vert 1 - h_s(1+\nu_s)/\ell_{c0}\vert$, as shown in Fig.~\ref{cs} (top right), with a crossover controlled by the thickness of the substrate $h_s$.
Using $F_T=F_A-k_c\Delta\ell$,  we solve for  both $\Delta\ell$ and $F_T$, shown in Fig.~\ref{cs} (bottom) as functions of the substrate thickness and stiffness.  For very thin ($h_s\rightarrow 0$) or infinitely rigid substrates, where the substrate elasticity becomes local,
$\Delta\ell=F_A/(k_c + k_a/2)$, corresponding to a spring $k_c$ in parallel with a series of two focal adhesions springs $k_a$. In this limit the traction  force saturates to $F_T=k_a F_A/(2k_c+k_a)$.
Conversely, for a very soft substrate with $E_s\rightarrow 0$, the contraction is maximal and given by $F_A/k_c$,
and $F_T\rightarrow 0$.
The  substrate thickness above which both cell contraction and traction force saturate is controlled by the cell size and the substrate elasticity, in qualitative agreement with experiments~\cite{Lin2010}.
%
%

\paragraph{Contractile Cell Layer.}
The continuum limit can be obtained by considering a multi-mer of $N=\left[L/l_{c0}\right]$ contractile elemental ``cells", connected by springs representing cell-cell interactions. The outcome is a set of coupled equations for a contractile elastic medium.
For a cell layer of thickness $h_c<<L$ (Fig.~\ref{cell_layer}), the force balance equation, averaged over the cell thickness, is
\begin{equation}
Y_a\left[u(x)-u^s(x)\right]=h_c\partial_x \sigma(x)\;,
\label{ux}
\end{equation}
where $Y_a=k_a/(L\ell_{c0})$ describes the effective strength of the focal adhesions, $u(x)$ is the displacement field of the cellular medium at $z=h_s$, and $\sigma$ is the thickness-averaged cellular stress tensor, $\sigma(x)={1/h_c}\int_{h_s}^{h_s+h_c}dz\ \sigma_{xx}(x,z)$, given by
$\sigma(x)=B_c \partial_x u  + \sigma_a$,
with $B_c$ the longitudinal elastic modulus of the cell layer. The one dimensional model presented here may be relevant to wound healing assays, where the cell layer is a strip with $y$-translational invariance.
Although we have neglected components of the cellular displacements  and spatial variations along $z$, the cell elastic constants are those of a three-dimensional cellular medium.
%
%
The active stress
$\sigma_a=F_A/(Lh_c)$  arises from  acto-myosin contractility~\cite{Kruse2005}. The substrate deformation at the surface is
\begin{equation}
u^s(x)=h_c\int dx' G(x-x')\partial'_x\sigma(x')\;,
\label{usx}
\end{equation}
with $G(x)$  the elastic Green's function of a substrate of infinite extent in $x$, occupying the region $0\leq z\leq h_s$, evaluated at $z=h_s$.
Eqs.~\eqref{ux}-\eqref{usx} can be reduced to  integro-differential equations for the cellular stress, as
\begin{equation}
\ell^2_a\partial_x^2\sigma+\sigma_a=\sigma-B_cLh_c\partial_x^2\int_{0}^{L}dx'G(|x-x'|)\sigma(x')\;.
\label{central-eq-1d}
\end{equation}
The length scale $\ell_a=\sqrt{B_ch_c/Y_a}$ controls spatial variations of cellular stresses induced by the stiffness of the focal adhesions.  It is  the size of a region where the areal elastic energy density $Y_a\ell_a^2$ associated with focal adhesions is of order of the areal elastic energy density $B_ch_c$ of the cell layer.
For a cell monolayer with $B_c=1 \unit{kPa}$, $h_c=0.1 \unit{\mu m}$, $L=100 \unit{\mu m}$,  $\ell_{c0}=10\unit{\mu m}$ and  $k_a=2.5 \unit{nN/\mu m}$~\cite{Balaban2001}, we get $\ell_a\simeq 6.3 \unit{\mu m}$, comparable to traction penetration length seen in experiments on stiff microposts~\cite{Saez2005,Saez2007}.
The second term on the right hand side of Eq.~\eqref{central-eq-1d} describes spatial variations in the cellular stress due to the (generally nonlocal) coupling to the substrate.
In the following we examine solutions to Eq.~\eqref{central-eq-1d}, considering various limiting cases for the substrate thickness and analyze the dependence of traction stresses on cell size, substrate stiffness and substrate depth.
The equation governing stress distribution in two dimensional cell layers is derived in the Supplemental material~\cite{SMprl12}.

\paragraph{Thin substrate.}
If the substrate's elastic response can be approximated as local, as it is the case for  $h_s<<L$  or for cells on micropillar arrays, the  Green's function is given by
$G(x) = \frac{2h_s(1+\nu_s)}{L E_s} \delta(x)$. Eq.~\eqref{central-eq-1d} can then be written as $\ell^2_p\partial^2_x\sigma+\sigma_a = \sigma$, where, $\ell_p=\sqrt{B_ch_c/Y_{\text{eff}}}$ and $Y_{\text{eff}}^{-1}=Y_a^{-1} + 2h_s(1+\nu_s)/E_s$ describes the combined action of the focal adhesions and the substrate, acting like two linear elastic components in series.
Assuming zero external stresses at the boundary, i.e., $\sigma(0)=\sigma(L)=0$, the internal stress profile  is $\sigma(x)=\sigma_a\left(1-\cosh{\left[(L-2x)/2\ell_{p}\right]}/\cosh{[L/2\ell_{p}]}\right)$~\cite{Edwards2011,Banerjee2011,Mertz2012}. The traction stress $T(x)=Y_{\text{eff}}u(x)$, is localized within a length $\ell_{p}$ from the edge of the cell layer.  The penetration length $\ell_{p}$ can be written as $\ell_{p}=\sqrt{\ell_a^2 + \ell_s^2}$, with $\ell_s=\sqrt{\frac{2B_c h_c h_s}{E_s/(1+\nu_s)}}$ the square root of the ratio of the cell's elastic energy to the elastic energy density of the substrate.  This form highlights the interplay of focal adhesion stiffness and substrate stiffness in controlling spatial variation of stresses in the lateral ($x$) direction. The two act like springs in series, where the weaker spring controls the response. If $Y_a<<\frac{E_s}{2(1+\nu-s)h_s}$, then $\ell_{p}\simeq\ell_a$ and the stiff substrate has no effect. Conversely, if the focal adhesions are stiffer than the substrate, then $\ell_{p}\simeq\ell_s$. For an elastic substrate with $h_s=10 \unit{\mu m}$,  $\nu_s=0.4$ and  $E_s$ in the range $0.01-100 \unit{k Pa}$, $\ell_s$ lies in the range $0.2-17 \unit{\mu m}$. This leads to typical values of $\ell_p$ in the range $6.3-18 \unit{\mu m}$ for a cell layer of length $100 \unit{\mu m}$, consistent with experimentally observed traction penetration lengths on thin continuous substrates~\cite{Mertz2012} and on micropillar posts~\cite{Saez2010}.

\paragraph{Infinitely thick substrate.}
If $h_s>>L$,
the substrate Green's function can be approximated as that of an elastic half plane,
$G(x)=-\frac{2}{\pi LE_s}\left[\gamma +\log{(\vert x \vert/L)}\right]$,
with $\gamma$ the Euler constant ~\cite{Barber2010}.
The solution of Eq. \eqref{central-eq-1d} with boundary conditions $\sigma(0)=\sigma(L)=0$ can be obtained by expanding $\sigma(x)$ in a Fourier sine series as, $\sigma(x)=\sum_{n=1}^{\infty} \sigma_n \sin{(n\pi x/L)}$
 and solving the
coupled algebraic equations for the Fourier amplitudes $\sigma_n$ given in the Supplementary Material~\cite{SMprl12}.
The effect of the nonlocal elasticity of the substrate is controlled by yet another length scale
$\ell_{s\infty}=\sqrt{\frac{4B_c h_cL}{\pi E_s}}$ that can be obtained from the length $\ell_s$ introduced in the case of thin substrate  by the replacement $h_s\rightarrow L$ and $(1+\nu_s)\rightarrow 2/\pi$. This highlights the known fact that cells or cell layers only ``feel" the substrate up to a thickness comparable to their lateral size $L$. For parameter values quoted in the preceding paragraphs, $\ell_{s\infty}$ takes values between $0.35-35 \unit{\mu m}$ for $E_s$ in the range $0.01-100 \unit{kPa}$, indicating that the thin/thick substrate crossover,
although not  observable in isolated cells,  should be seen experimentally in cohesive cell layers where the lateral extent can exceed $100 \unit{\mu m}$. The cellular stress and substrate displacement profiles obtained numerically by summing the Fourier series are shown in Fig.~\ref{subs} (top). The lateral variation of stresses is now controlled  by the length scale $\ell_p=\sqrt{\ell_a^2+\ell_{s\infty}^2}$. One consequence of nonlocal substrate elasticity  is that the substrate deformation shown in the top right frame of Fig.~\ref{subs} extends outside the region occupied by the cell layer, indicated by the two vertical dashed lines. The profile of the local traction stress displayed in Fig.~\ref{subs} (bottom frame) shows that the traction stress is localized near the edge of the cell layer and its magnitude increases with substrate stiffness.
The inset to Fig.~\ref{subs} (bottom right) shows the magnitude of the net contractile moment defined as ${\cal P}=\int_{-\infty}^\infty dx x T(x)$. This quantity is negative, as expected for contractile systems.
Its magnitude increases with $E_s$ at a rate consistent with experiments, with a $25\%$ rise in $|{\cal P}|$ upon increasing the substrate stiffness by $40\%$~\cite{Wang2002}, and saturates for very stiff substrates.
%
%
%

\paragraph{Substrate of Finite Thickness.}
Finally, we consider a substrate of finite thickness, $h_s$.
The calculations are carried out using the approximate Green's function given in Eq.~\eqref{G-finite}, with the replacement $\ell_{c0}\rightarrow L$.
The variation of the net contractile moment with $h_s$ for $E_s=10 \unit{Pa}$ is shown in Fig.~\ref{subs} (bottom left inset). As seen previously in experiments~\cite{Lin2010}, $\vert{\cal P}\vert$ drops sharply with increasing substrate thickness, quickly reaching the asymptotic value corresponding to infinitely thick substrates. Thinner substrates are effectively stiffer than thick ones, inducing larger contractile moments.
Our analysis suggests a general expression for the penetration length $\ell_p$ that interpolates between the thin and thick substrates limits,
\begin{equation}
\label{lp}
\ell_p=\sqrt{\frac{B_ch_c}{Y_a}+\frac{B_ch_c}{\pi E_s}h_{\rm eff}}\;.
\end{equation}
Stress penetration is controlled by a substrate layer of effective thickness $h_{\rm eff}^{-1}=\frac{1}{h_s2\pi(1+\nu_s)}+\frac{1}{L}$ given by the geometric mean of the actual substrate thickness $h_s$ and the lateral dimension $L$ of the cell or cell layer. If $h_s<<L$, then $h_{\rm eff}\approx 2\pi h_s(1+\nu_s)$ and stress penetration is not affected by cell layer size, as in the experiments of \cite{Mertz2012}. On the other hand, if $h_s>>L$, then cells only feel the effect of the substrate down to an effective depth $L$.
\paragraph{Discussion.}
In summary, we have examined the dependence of traction stresses in adhering cell layers on the mechanical and geometrical properties of the substrate. Using a generic non-local model, we provide analytical results for the effect of cell and substrate properties on the stress penetration length, that can be tested in experiments. Although the analysis presented here is restricted to one dimensional layers, isotropic planar cell layers with spherical symmetry can also be considered analytically~\cite{Banerjee2012}, with similar predictions for the dependence of traction fields and their moments on substrate mechanical and geometrical properties. The scaling of traction moments on cell layer size is, however, different in two dimensions~\cite{Mertz2012}. The model can be extended to incorporate the effects of cell polarization, spatial variations in contractility, heterogeneities in the cell layer or anisotropic elasticity of the substrate.

We thank Eric Dufresne and Aaron Mertz for many useful discussions and the anonymous referees for valuable comments. This work was supported by the National Science Foundation through awards DMR-0806511,  DMR-1004789 and DGE-1068780.

\bibliography{cell-substrate}

\end{document}